\documentclass[aps,preprint,aps]{revtex4}%
\usepackage{graphicx}
\usepackage{amsmath}
\usepackage{amsfonts}
\usepackage{amssymb}%
\begin{document}

\title{ Magneto-optical properties of (Ga,Mn)As:\\ an ab--initio \textbf{determination}}
\author{Alessandro Stroppa}\affiliation{Faculty of Physics, University of Vienna, and Center for
Computational Materials Science, Universit\"at Wien, Sensengasse
8/12, A-1090 Wien, Austria} \email{alessandro.stroppa@univie.ac.at}
\author{Silvia Picozzi}\affiliation{Consiglio Nazionale delle Ricerche, Istituto Nazionale Fisica della Materia (CNR-INFM),  \\
CASTI Regional Lab., 67010 Coppito (L'Aquila), Italy}
\author{Alessandra Continenza} \affiliation{CNISM - Dipartimento di Fisica,
Universit\`a degli Studi dell'Aquila,\\ 67010 Coppito (L'Aquila),
Italy }
\author{MiYoung Kim} \affiliation{BK21 Frontier Physics Research Division, Seoul National University, Seoul,\\ 151-747 (Korea)}
\author{Arthur J. Freeman}\affiliation{Department of Physics and Astronomy, Northwestern University, Evanston,\\ IL 60208 (U.S.A.)}
\date{\today}

\begin{abstract}
The magneto-optical properties of (Ga,Mn)As have been determined
within density functional theory using the highly precise
full-potential linear augmented plane wave (FLAPW) method. A
detailed investigation of the electronic and magnetic properties in
connection to the magneto-optic effects is reported. The spectral
features of the optical tensor in the 0-10 eV energy range are
analyzed in terms of the band structure and density of states and
the essential role of the dipole matrix elements is highlighted by
means of Brillouin zone \emph{dissection}. Using an explicit
representation of the Kerr angle in terms of real and imaginary parts
of the tensor components, a careful analysis of the Kerr spectra is
also presented. The results of our study can be summarized as
follows: i) different types of interband transitions do contribute
in shaping the conductivity tensor; ii) the dipole matrix elements
are important in obtaining the correct optical spectra; iii)
different regions in the irreducible Brillouin zone contribute to
the conductivity very differently; iv) a minimum in the Re
$\sigma_{xx}$ spectra \emph{can} give rise to a large Kerr rotation
angle in the same energy region; and v) materials engineering
 via the  \emph{magneto-optical Kerr effect} is possible
provided that the
 electronic structure of the material can be tuned in
 such a way as to \emph{enhance} the depth of the minima of
 Re $\sigma_{xx}$.
\end{abstract}

\maketitle

\section{Introduction}
  The magneto-optical Kerr effect (MOKE),  discovered in 1877,
\cite{kerr} consists in the rotation  of the polarization plane of
linearly polarized light with respect to that of the incident light
reflected from a magnetic solid surface. But only very recently has
this property become the subject of intense investigations.
 The reasons are twofold. First,
 one can exploit this effect to read suitably
magnetically stored  information using optical means in modern high
density data storage technology, erasable video and audio disks
(magneto-optical disks).\cite{kerrdisk01,kerrdisk02,Kerrdisk1,Kerrdisk2,Kerrdisk3} Second, MOKE is now regarded 
as a powerful probe in many fields of research, such as a microscopy
for domain observation,\cite{microscopy1} surface magnetism,
magnetic interlayer coupling in multilayers, plasma resonance
effects in thin layers,  and structural  and magnetic
anisotropies.\cite{Freeman1,microscopy2} Magneto-optical
measurements are also a valuable tool in the study of the magnetic
properties and electronic structure of magnetic materials. Further,
it is well known that optical reflection measurements can be used to
determine the diagonal elements of the dielectric tensor. However,
the reflection spectra of many intermetallic compounds do not show a
pronounced  fine structure and so the information  on the
electronic structure obtained through reflection spectra is not very
detailed. In contrast, the Kerr effect shows a finer structure and
gives important insights into the properties of transition-metal and
rare-earth compounds.\cite{kerrinfo1,kerrinfo2,kerrinfo3}

In the last ten years, transition metal doped semiconductors have
attracted considerable attention in the  field of
\emph{spintronics},\cite{spintronic1,spintronic2,spintronic3} where
Ga$_{1-x}$Mn$_{x}$As is  by far the most studied material suitable
for semiconductor-based spintronic devices.\cite{spin1,spin2,spin3}
Interestingly, the giant magneto-optical effects observed in some of
these systems are especially attractive for optical
applications.\cite{huge1,huge2,huge3} For instance, optical
isolators based on magnetic semiconductors might be ideal components
for high speed optical communication systems.\cite{MagnetoOptics}
Furthermore, magneto-optical measurements have been playing an
important role in clarifying the exchange interaction and the
electronic structure of these ferromagnetic semiconductors, as they
provide very detailed information on the influence of broken
time-reversal symmetry in itinerant electron quasi-particle
states.\cite{MagnetoOptics}

Despite the increasing role of magnetic semiconductors over  the
last years, and the extensive studies available on their electronic
 and magnetic properties,\cite{ab1,ab2,ab3,ab4,ab5,ab6,ab7} only
very recently have \emph{ab--initio} calculations of their
magneto-optical properties been performed.\cite{Silvia,Weng,Weng2}
Weng \emph{et al.} studied the electronic structure and polar
magneto-optical Kerr effect of transition metal chalcogenides, such
as CrSe, CrTe, and  VTe in zinc-blende and wurtzite
structures, by full-potential density-functional
calculations\cite{Weng}, while in a following paper Weng \emph{et
al.} analyzed the magneto-optical response of Zn$_{1-x}$Cr$_x$Te
ordered alloys. In Ref.~\onlinecite{Silvia}, some of us investigated
the magneto-optical properties of Ga$_{1-x}$Mn$_{x}$As with highly
precise first-principles density-functional FLAPW calculations in
order to perform an accurate comparison with experiments, focusing
on the effect of Mn concentration and of the occupied sites.

The purpose of the present work is to go beyond the analysis
presented in Ref.\ \onlinecite{Silvia}, by performing an extensive
study focused on the optical conductivity tensor as well as on the
Kerr spectra. Here we mainly focus on (Ga,Mn)As ferromagnetic
semiconductors in the high concentration limit by considering a 25\%
concentration of Mn substituting for Ga. Although this can be
considered as a rather unrealistic case due to the low solubility of
Mn in the host GaAs crystal, important aspects of the interplay
between the underlying electronic structure and the MO Kerr effect
(MOKE)
  can be gained and then extended to the low-concentration limit. 
  We therefore studied the optical conductivity
  in connection with the  electronic structure. As analysis tools, we
  considered the  density of states
(DOS) and band structure in order to elucidate  the origin of the
features of the optical spectra in terms of electronic transitions.
The origin of these features are further investigated by performing
an analysis of the electronic transitions throughout the Brillouin
zone (\textbf{k}-space dissection). In this way, the contributions
coming from the different regions of the Brillouin zone can be
separated out and the role of the dipole matrix elements can be
quantitatively analyzed. Finally, in order to compare our results
with available experiments,
 we  considered a more dilute
case, namely  $x$=6.25 \% .

After the analysis of the optical properties, we turn our attention
to the study of the Kerr spectra. Our approach is based on a
\emph{real} representation of the Kerr angle. We show, using such a
representation, that useful analytical relations can be derived and
can be used to gain insights into the microscopic quantities that
determine the magnitude and frequency position of the main features
in the Kerr rotation spectrum.

The work is organized as follows: in the next section, we briefly
describe the theoretical methods and provide computational details;
in Sect.\ \ref{Elemagne}, we review the electronic and magnetic
properties of  Ga$_{0.75}$Mn$_{0.25}$As; Sect.\ \ref{optic} is
devoted to the analysis of the optical conductivity tensor in terms
of the band structure; in Sect.\ \ref{k-section} we perform a full
$\vec{k}$-space analysis in order to highlight the role of dipole
matrix elements in shaping the optical spectra; in Sect.\
\ref{optic-magneto}, we analyze in detail the Kerr spectra; finally,
in Sect.\ \ref{conclu},
we draw our conclusions.\\

\section{Theoretical framework and computational details}
In this work, the Kohn-Sham equations are solved self-consistently,
using the full--potential linearized augmented
 plane wave (FLAPW) method.\cite{FLAPW} We used the local spin density approximation (LSDA) for
 the exchange-correlation functional, with the parametrization of Hedin-Lundqvist.\cite{LDAHL}
  The spin-orbit coupling (SOC)
is essential to obtain the orbital magnetic moment as well as the
magneto-optical effects:\cite{spinorbit1,spinorbit2} in the
evaluation of the optical conductivity, the spin orbit effect is
neglected in the self--consistent iterations but is included in a
second variational step.

The optical conductivity tensor is calculated according
 to the Kubo formula in the linear
 response theory:\cite{kubo1,kubo2,kubo3}
\begin{equation}\label{KUBO}
 \sigma_{\alpha,\beta}(\omega)=\frac{Ve^{2}}{8\pi^{2}\hbar
    m^{2}\omega}\sum_{n,n^{'}}\int
    d^{3}\vec{k}<\vec{k}n|p_{\alpha}|\vec{k}n^{'}><\vec{k}n^{'}|p_{\beta}|\vec{k}n>f_{\vec{k}n}(1-f_{\vec{k}n^{'}})\delta(\epsilon_{\vec{k}n^{'}}-
    \epsilon_{\vec{k}n}-\hbar\omega)
\end{equation}
where $\alpha,\beta$=1, 2, 3; $p_{\alpha}$ are components of the
momentum operator, f$_{\vec{k}n}$ is the Fermi distribution function
ensuring that only transitions from occupied to unoccupied states
are considered; $|\vec{k}n>$ is the crystal wave function,
corresponding to the  Kohn-Sham eigenvalue $\epsilon_{\vec{k}n}$
with crystal momentum $\vec{k}$; and the delta function warrants
total energy conservation. The above formula considers infinite
lifetime of excited Bloch electronic states. In order to take into
account finite lifetime effects, we broadened the optical spectra by
fixing the interband relaxation time for excited states to 0.3 eV.

Equation \ref{KUBO} contains a double sum over all energy bands,
which naturally splits into the so-called interband contributions,
\emph{i.e.}, $n\neq n^{'}$, and the intraband contributions, $n=n^{'}$,
that is:
\begin{equation}\label{interintra}
\sigma_{\alpha,\beta}=\sigma_{\alpha\beta}^{inter}(\omega)+\sigma_{\alpha,\beta}^{intra}(\omega)
\end{equation} For the diagonal tensor components, both terms are important and
 should be considered simultaneously.
For metals (or half-metals), the intraband contribution to the
diagonal component of $\sigma$ is usually described  by
the phenomenological expression according to the Drude-Sommerfeld model:\cite{drude1,drude2}\\
\begin{equation}\label{drude_form}
    \sigma_{D}(\omega)=\frac{\omega_{P}^{2}}{4\pi[(\frac{1}{\tau_{1}})-i\omega]}
\end{equation}

Within the Drude theory framework,  the complex conductivity is
fully characterized by two parameters: the plasma frequency
$\omega_{P}$ and the relaxation rate $\gamma_{1}=1/\tau_{1}$.
 The intraband relaxation time, $\tau_{1}$, characterizes the scattering of
charge carriers,   which depends on the amount of defects and therefore varies from sample to sample.
Here we choose a
fixed value of 0.7 eV for $\gamma_{1}$ in all our simulations.   It is worth noting that this 
 value does not affect our discussion below, and, in any case, variations of  its value
   between  0.2 and 0.7 eV, which include the usual variability range for
 $\gamma_{1},$\cite{Gamma1} 
  lead to a
negligible effect on the optical conductivity for energy values
larger than 1 eV. The unscreened plasma frequency is obtained
integrating over the Fermi surface:

\begin{equation}\label{wp}
    \omega_{P,ii}^{2}=\frac{8\pi e^{2}}{V}\sum_{\textbf{k}n,s}|<\textbf{k}ns|p_{i}|\textbf{k}ns>|^{2}
    \delta(\varepsilon_{\textbf{k}n}-\varepsilon_{F})
\end{equation}
where $V$ is the volume of the primitive cell, $\varepsilon_{F}$ the
Fermi energy, $e$   the electron charge and $s$ is the spin. The
calculated value of $\omega_{P}$ is 2.42$\times$10$^{14}$ Hz (2.75
eV). The intraband contribution to the off-diagonal optical
conductivity is very small  and is usually  neglected in the
case of magnetically ordered materials.\cite{Antonov}

In the present work, we consider the Kerr effect in the so-called
polar geometry, where the magnetization vector is oriented
perpendicular  to the reflective surface and parallel to the plane
of incidence. In such a case, the Kerr rotation angle
 $\theta_{k}(\omega)$ and its ellipticity $\eta_{k}(\omega)$ can be obtained
 from the conductivity tensor as:\\
 \begin{eqnarray}\label{Kerr}
  \theta_{k}(\omega)+i \eta_{k}(\omega)&=& -\frac{\sigma_{xy}(\omega)}{\sigma_{xx}
  (\omega)\sqrt{1+i(4\pi/\omega)\sigma_{xx}(\omega)}}
 \end{eqnarray}.

Our study is based on the supercell approach where one of the Ga
atoms in an 8-atom cell of zinc-blende GaAs is replaced by a Mn
atom, thus simulating an ordered alloy with an x=25\% Mn
concentration. The lattice parameter is chosen equal to the
experimental lattice parameter of GaAs (5.65 \AA).  For larger
unit-cells, {\em i.e.} x = 6.25 \%, we refer to the work of Picozzi
{\em et al.},\cite{Silvia} where the pertinent computational and
structural details are given.

As far as the other technical details are concerned, we used a
wave-vector  cutoff of the basis set equal to
 $K_{max}$=3.5 a.u. and an angular momentum expansion up to
 $l_{max}=8$ for both the potential and charge density. The
 muffin-tin radius, $R_{MT}$, for Mn, Ga and As were chosen equal to
 2.1, 2.3, and 2.1 a.u., respectively.
 The relaxed internal atomic positions were obtained by
total-energy and atomic-force minimization using the
Hellmann-Feynman theorem\cite{Feynman} with residual forces
 below 0.015 eV/\AA. The sampling of the
irreducible wedge of the
 Brillouin zone (IBZ) was performed using the special $k$-point
 method.\cite{MP} In order to speed up the convergence of the
 sampling, each eigenvalue is smeared with a Gaussian function of
 width 0.03 eV . The ground-state electronic structure is calculated
 using  a 4$\times$4$\times$4 cubic mesh.
 We  checked that these computational
 parameters are accurate enough to obtain total energies and magnetic
 moments within 10-15 meV/Mn and 0.01 $\mu_{B}$ (keeping the
 muffin-tin radii fixed), respectively.
 On the other hand, in order to accurately compute
 the optical conductivity, the $k$ integration must be carefully taken care of. We
   investigated its accuracy by varying the number of $k$
   points up to a (10,10,10) shell (35 $k$-points in the IBZ):
we get the Kerr rotation angle and ellipticity within a few
hundredths of a degree already using a (4,4,4) shell.
\\

\section{Electronic and magnetic properties}\label{Elemagne}

Before discussing the magneto-optical (MO) properties, we consider
in some detail the electronic structure and the magnetic properties
of Ga$_{0.75}$Mn$_{0.25}$As. Although shown and discussed in several
previous
publications,\cite{GaMnAs1,GaMnAs2,GaMnAs3,GaMnAs4,previous1,previous2,previous3,previous4}
we show them  in Fig.~\ref{dos} and Fig.~\ref{bands}, as it is
fundamental to have in mind a clear picture of the electronic and
magnetic properties of the compound considered. This is needed in
order to consistently follow the discussion of the optical
conductivity tensor (see below), which is the main issue of the
present work. In particular, in Fig.~\ref{dos} we show the spin
polarized total density of states (DOS) and the site angular
momentum-projected DOS (PDOS)  for Mn and nearest- and
next-nearest-neighboring atoms, that is As and Ga atoms,
respectively. In Fig.~\ref{bands}, we show the spin-polarized band
structure (not including spin-orbit coupling), along the
$\Gamma$-X-M-$\Gamma$-R-M symmetry directions, for the majority
(left) and minority (right) bands. Numbers in round brackets label
groups of bands (eventually degenerate) at the $\Gamma$ point.
Vertical dashed lines denote the electronic transitions giving rise
to spectral features in Re $\sigma_{xx}$ (see next Section), and in
the following, we discuss both Figs.~\ref{dos} and ~\ref{bands}. The
total DOS (see Fig.\ref{dos}, topmost panel) shows the \emph{nearly}
half-metallic (n-HM) nature of the electronic ground state: the
majority bands cross the Fermi level (set to zero) whereas a net gap
opens in the minority bands with the Fermi level cutting the
minority conduction band near $\Gamma$.

The total DOS curve is composed of several characteristic structures
whose origin can be easily recognized by looking at the partial
(P)DOS curves. In the following, we only consider the upper part of
the valence band (above the ionicity gap, that is above $-$7.5 eV)
and the lower part of the conduction band (up to $\sim$ 5 eV) which
define the energy window relevant for the optical transitions we are
interested in.
 The  peak centered at $-$6 eV in the DOS curve
 (for both spin components) is dominated by the energetically
 low-lying  Ga-$s$ states, and corresponds to the lowest lying bands in Fig.~\ref{bands}, [(1) and
(1a)], which are rather (surprisingly) well dispersed  along
$\Gamma$-X-M-$\Gamma$, and almost flat in the $\Gamma$-R-M
direction. As the energy increases, there is a peak at $-$5.0
($-$4.6) eV in the majority (minority) component, which mainly
derives  from $s$-Mn; it corresponds to the next low-lying band
shown in Fig.~\ref{bands}, [(2) and (2a)], which is degenerate with
the lowest band [(1) and (1a)] at $R$.

A broader structure in the total DOS extends from about $-$4.0 eV up
to $E_{F}$ (majority component) and up to about $-$1.0 eV (minority
component). Let us consider the majority component. The sharp peak
at $-$3.0 eV originates from Mn-$d$ orbitals and corresponds to the
group of bands in Fig.~\ref{bands} [(3)]; these bands are degenerate
at $\Gamma$ and are split into three bands along $\Gamma$-M and into
two bands along $\Gamma$-R. Inspection of the eigenvalue
decomposition shows that these states have $t_{2}$ symmetry at
$\Gamma$ and thus they hybridize with anion $p$ states (see the
corresponding PDOS curve in Fig.~\ref{dos}). In general, the higher
energy bands [groups (4) and (5) in Fig.~\ref{bands}] mostly derive
from hybridization between Mn-$d$ and As-$p$ states, as well as the bands [(6)
and (7)] around $E_{F}$. 
 The
Mn-$s$ states contribute little to the defect bands since they
obviously lie lower in energy than the Mn-$d$ and -$p$ states. One
also recognizes the intra-atomic hybridization (among states of the
same atom) occurring for Mn-$p$ and -$d$ states: the corresponding
DOS curves overlap in a wide energy range. Higher bands [(8)]  are
mostly derived from cation and anion states. For the minority
component, the valence bands, [(3a), (4a) and (5a) in
Fig.~\ref{bands}], are mostly dominated by As-$p$ states. The Mn-$d$
states are pushed upward in energy above $E_{F}$, due to exchange
splitting: these states are almost unoccupied and remain fairly
narrow in energy above the gap. The conduction bands [(6a), (7a)
and (8a)] are composed of Mn-$d$ and -$p$ states just above $E_{F}$,
while higher bands [(9a)] mainly by Ga-$s$ and  As-$p$ states.
The same  Mn-$p$,-$d$ intra-atomic mixing is found in the minority
component above $E_{F}$ (the Mn-$p$ and -$d$ DOS are localized in
the same energy range).

\section{Optical conductivity}\label{optic}

\subsection{General considerations}
 In Fig. \ref{sigmaxx}, we show the (a) real (left) and (b) 
 imaginary (right) parts
  of the diagonal (top) and off-diagonal
(bottom) components of the conductivity, in the range 0-10 eV. For
simplicity, we did not include the Drude term in the conductivity
curves shown. In the real part of the diagonal component (Re
$\sigma_{xx}$), we highlight some features with vertical lines,
which will be discussed in closer detail later.

The Re $\sigma_{xx}$ curve  has a quite broad peak at around 3.8 eV
whereas the Im $\sigma_{xx}$ curve shows  a deep minimum in the low
energy range (around 2.3 eV) and a shoulder above 6.8 eV.
 The  Re $\sigma_{xy}$ curve shows small amplitude oscillations in the energy
range 0$-$4 eV; a deep minimum is found at $\sim$ 5 eV, and a
maximum follows at $\sim$ 6.8 eV followed again by oscillations with
small amplitude.
 The imaginary component, Im $\sigma_{xy}$,  has  similar characteristics in the low
 and high energy range (small amplitude oscillations),
 but a maximum develops at $\sim$ 4.6 eV and a minimum  follows at $\sim$ 6.4 eV.

The absorptive part of the off-diagonal optical conductivity, Im
$\sigma_{xy}$, has a direct physical interpretation. It is
proportional to the difference in absorption rate of the left and
right circularly polarized (LCP and RCP) light, and its sign is
directly related to the spin polarization of the states responsible
for the interband transitions producing the structures in the
spectrum.\cite{Ebert} The  peak in Im $\sigma_{xy}$ around 4.6 eV
suggests that interband transitions related to LCP light should be
stronger in this energy region. On the other hand, around 6.4 eV, Im
$\sigma_{xy}$ shows a deep minimum value, indicating the dominance
of  interband transitions related to RCP. Finally, Im $\sigma_{xy}$
is negligible when the absorption coefficient for RCP is  equal to
that of LCP - i.e., in the lowest and highest parts of the spectrum.

\subsection{Analysis in terms of band structure}
Since Re $\sigma_{xx}$ is directly related to the density of states
and transition probabilities,\cite{Ebert}  we investigate the
calculated energy band structure in order to analyze the details of
the spectra. It has been known for quite some time that both the
spin-orbit (SO) interaction and the exchange splitting are needed to
produce a non-zero Kerr effect.\cite{spinorbit1,Kerr2,Kerr3}
Therefore, a proper analysis should be performed considering the
band structure with spin-orbit (SO) included. However, we remark
that the calculated band structure with SO (not shown here) is quite
similar to that formed by the superposition of the majority and
minority spin bands (without SO).  This corresponds to a spin-orbit
coupling interaction that is small compared to the exchange
splitting, so we can qualitatively consider the
former as a perturbation to the latter:\cite{spin-exchange} this
allows us to restrict our discussion to the usual spin-polarized
band structure, thus neglecting the mixing   of majority and
minority spin components. In this way, a decomposition with respect
to each separate spin channel (spin up, down) can be done; this
decomposition is almost uniquely determined if the spin-orbit
coupling is weak. Inspection of the eigenvalue
decomposition at $\Gamma$ shows that the exchange-splitting 
of As-$p$ and Mn-$d_{e_{g}}$  states are
$\sim$ 0.03 and 2.5 eV, respectively, whereas the 
spin-orbit splitting is $\sim$ 0.02 and 0.002 eV, 
respectively. In this case, the main effect of the SO interaction is just to remove
some accidental and systematic degeneracies\cite{accidental,Spin-orbitDeg,Refsystematic}.

In Fig. \ref{bands}, we can recognize three different contributions
coming from the band structure
 which build up the optical spectra: (a) close parallel bands
crossing $E_{F}$; (b) bands which are degenerate (or almost
degenerate) at one point and then separate out,  moving below and/or
above $E_{F}$; and (c) occupied/unoccupied bands which are almost
parallel in a significant part of the Brillouin zone.\cite{parallel}
The first type does contribute to the conductivity in the low energy
range and gives rise to bumps of interband transitions; the second
type gives rise to an almost constant amount of interband
transitions with a distinct onset energy; the third type gives rise
to peaks at higher energies. In the following, we refer to
\emph{bands} although a more appropriate discussion should be done
in terms of \emph{constant eigenvalue surfaces} in reciprocal
space.\cite{surface} However, the purpose of this section is to
analyze on a qualitative level the origin of the different
structures in Re $\sigma_{xx}$. For this purpose, we highlight the
relevant electronic transitions (see Fig. \ref{bands}) that give
rise to the spectral features in the Re $\sigma_{xx}$ curve at 0.4,
1.1, 2.8 and 3.8 eV (see Fig. \ref{sigmaxx}), respectively.

Re $\sigma_{xx}$ is non-negligible already at rather low energies
(as low as 0.5 eV, see Fig.~\ref{sigmaxx}). As can be seen from
Fig.~\ref{bands}, interband
 transitions are available at this energy
 due to bands which are degenerate at some $\vec{k}$ points at
 energies slightly below (or above) $E_{F}$ and then separate out, moving above $E_{F}$ [type (b) transitions]. For
 example, in the minority component,
  the main contribution in
 this energy range comes from the two bands [(6a)] below $E_{F}$, which are degenerate at $\Gamma$ and split along
$\Gamma$-X and $\Gamma$-M. Incidentally, we note that the same bands
are responsible for the n-HM character of the (Ga,Mn)As compound. At
very low energy, the bands along $\Gamma$-M do contribute to the
conductivity, whereas those along $\Gamma$-X do so at higher
energies.

Consider in detail the minority bands along $\Gamma$-X with energies
just below $E_{F}$ [bands (6a) at $\Gamma$, see Fig. \ref{bands}]. A
careful inspection of the eigenvalue decomposition into
atomic-site-projected wave functions shows that these bands have a
prevalent $d$ character but with a non-negligible contribution from
anion $p$  as well as $s$ cation states. Moving away from the zone
center, the $d$ as well as $p$ contributions increase in  the higher
energies of these bands, while the $s$
 contribution decreases; the opposite occurs
 for the lower band.
The $d-p$ and $s-p$ mixing is crucial for the onset of the  dipole
transitions among these states at very low excitation energies which
otherwise would be forbidden by selection rules. Moving away from
$\Gamma$, their energy separation increases, as is typical for  type
(b); as the lower band crosses $E_{F}$, the interband transitions
stop contributing at an excitation energy around 0.40
eV. This  corresponds to the first flat region
in Re $\sigma_{xx}$ starting around 0.40 eV. Other electronic
transitions at 0.40 eV are shown in Fig. \ref{bands} with lines near
the M point in the spin-up bands.

 Upon further increase of the excitation
energy, additional type (b) structures start  contributing, like the
majority group of bands [(7)] above $E_{F}$ and degenerate at R which
split along R-$\Gamma$ and R-M. In particular, along R-$\Gamma$, one
band remains above $E_{F}$, while the other goes below. This causes
an increase of Re $\sigma_{xx}$, at energies lower than 1.0 eV. In
this energy range, type (a) bands also contribute, like the almost
parallel majority bands which cross $E_{F}$ along $\Gamma$-M, and
those crossing $E_{F}$ along $\Gamma$-X [all of them deriving from
bands (6) and (7)].

As the photon energy reaches $\sim$ 1.1 eV, several transitions are
activated,  accounting  for the bump in the  conductivity. The
transitions at 1.1 eV are shown with blue (dotted) lines in Fig.
\ref{bands}. In the majority (minority) component, they are mainly
localized near the R ($\Gamma$) point [mostly type (c) transitions].

At $\sim$ 2.8 eV,  we have a bump in the Re $\sigma_{xx}$ (see Fig.
\ref{sigmaxx}), First, let's consider the minority component. The
almost dispersionless occupied band around $-$1.0 eV [(5a)] is
coupled to the group of bands around 1.5 eV [(8a)]. These are type
(c) transitions and they give rise to the bump in Re $\sigma_{xx}$
around 2.8 eV. The transitions which match the energy of 2.8 eV
are shown by red (dashed) lines in Fig. \ref{bands}.
In the majority component there is also a contribution coming from
the group of bands (5) and (7).  As the photon energy increases, Re
$\sigma_{xx}$ reaches its maximum value, at around 3.8 eV.
There are several contributions to this peak: they involve
degenerate minority bands at high symmetry points, like the groups
of bands at $\Gamma$ centered at $-$2.5 eV [(4a)] and 1.5 eV [(8a)]
and at R with  energies $\sim$ $-$2.0 eV and 2.0 eV, respectively.
In the majority components, more type (c) transitions contribute to
the maximum of Re $\sigma_{xx}$. They mainly involve groups of
degenerate bands at $\Gamma$ (occupied/unoccupied) which split along
$\Gamma$-M: they are not flat along this symmetry line but are
  almost parallel near $\Gamma$. For instance, the highest of this group of bands [(5)]
split along $\Gamma$-M couples with the lowest of the group of bands
[(8)].

As the  photon energy increases, the lowest majority and minority
energy bands between $-$5 and $-$7 eV below $E_{F}$ come into play.
 The dispersion of the
occupied as well as the unoccupied bands involved in the transitions
increases, and this correlates with   the 
conductivity decrease.

It can be useful to consider the role of the electrons with spin up
and spin down in shaping the spectra of Re $\sigma_{xx}$. Again,
this  can be  done  on a qualitative level only. In fact, the dipole
matrix elements enter quadratically in the calculation of the
conductivity (see Eq.~\ref{KUBO}) and \emph{interference effects}
can be lost by separating the spin-up or spin-down contributions to
the conductivity. Indeed, from Eq. \ref{KUBO}, we have, for
$\alpha$=$\beta$=$x$ and neglecting the $\vec{k}$
index:\\

\begin{equation}
< n|p_{x}| n^{'}>< n^{'}|p_{x}| n>=|< n|p_{x}| n^{'}>|^{2}
\end{equation}

where $n$ includes the spin indices. Expressing the spin indices,
and neglecting spin-flip transitions\cite{Spintransition,SpinFlip} we have:

\begin{equation}\label{Interference}
|< n|p_{x}|
n^{'}>|^{2}=|\sum_{ss^{'}}p^{ss^{'}}_{x,nn^{'}}|^{2}=|p^{\uparrow\uparrow}_{x,nn^{'}}|^{2}+
|p^{\downarrow\downarrow}_{x,nn^{'}}|^{2}+
2Re(p^{\uparrow\uparrow}_{x,nn^{'}}p^{\downarrow\downarrow
*}_{x,nn^{'}})
\end{equation}

 In Fig.~\ref{joint}, we show the two spin
contributions to Re $\sigma_{xx}$, \emph{i.e.} neglecting the last term in Eq.
\ref{Interference}, in the range 0-10 eV.
Incidentally, we note that the \emph{sum} of the two separate contributions
(which does not include the interference effects coming from the last term in Eq.
\ref{Interference}), not shown in Fig.~\ref{joint}, differs in some fine details from
the curve shown in Fig.~\ref{sigmaxx} (which naturally includes the 
interference effects). This
confirms a posteriori that  the loss of
interference effects  due to neglecting the last term in Eq.~\ref{Interference}  introduces 
a negligible error and allows us to treat the two spin contributions to the 
conductivity separately. The two curves are qualitatively similar, and
they clearly show the origin of the bump at 2.8 eV and the maximum
at 3.8 eV: the former is mainly due to transitions involving
minority while the latter the majority spin bands. From  Fig.
\ref{bands} one could conclude that both the features at 2.8 and 3.8
eV arise from minority spin band transitions: this  is not exactly
true due to the role of the dipole matrix elements.

In the same Fig.~\ref{joint}, we also show the joint density of
states (JDOS), which can be derived from Eq.~\ref{KUBO} by replacing
the matrix elements with a constant factor set to 1 in our
case.\cite{Ebert} This quantity clearly  highlights the role of the
dipole matrix elements: as a matter of fact, the JDOS does not show
any relevant feature in the spectra stressing that the spectral
 features  observed occur primarily
 due to the very large matrix elements rather than to  a large joint
 density of states effect. On the same basis, the large shoulder above 6
 eV in the JDOS does not produce any peak in the conductivity
 spectra due to the small matrix elements.

\subsection{Comparison with experiments}
Finally, we compare our calculations with some experimental results.
For this purpose, we consider a lower concentration $x$ = 6.25$\%$
(1 Mn in a 32-atom cell) that  is closer to available
experiments.\cite{singley,burch,kojima}  In Fig.~\ref{optical} (a)
we show the  calculated real component of the
 diagonal  optical conductivity,
 Re $\sigma_{xx}$, as a function of the photon frequency, compared with
 spectra obtained from experiments and model calculations \cite{hanki} for
 Ga$_{0.95}$Mn$_{0.05}$As (at hole
concentration $p \sim$ 0.8 nm$^{-3}$), in the [0.01-2] eV energy
range. In this interval, our calculated conductivity is of the order
of 1-2$\cdot$10$^3$ (ohm$\cdot$cm)$^{-1}$, i.e., remarkably larger
(almost by an order of magnitude) than the experimental
values\cite{singley} but of the same order of magnitude as the
results of model calculations.\cite{hanki}
 The disagreement with experiments may be fully explained in terms of the suggested values of carrier
concentration in the experimental samples: $p \sim$ 0.35 nm$^{-3}$
for a Mn concentration on the order of $\sim$6$\%$, with a degree of
compensation on the order of 70-80$\%$.

In our systems, a pretty naive evaluation of the hole concentration
can be obtained by supposing that each Mn - with a nominal valence
of +2 - substituting a Ga atom gives rise to a hole; therefore, 1 Mn
atom in a 32 atom cell produces  a hole concentration of 1.38
nm$^{-3}$. It has been shown\cite{hanki} that the hole concentration
has
 a relevant influence on the optical conductivity: for example,
at very low energies ($<$0.01 eV) $\sigma_{xx}$ ranged \cite{hanki}
from $\sim$ 1.5x10$^2$  to $\sim$ 3.5x10$^3$ (ohm$\cdot$cm)$^{-1}$
for hole concentrations varying from  $p \sim$ 0.03 to 0.80
nm$^{-3}$. These values are very consistent with the range in which
our conductivity falls. Looking at the trend as a function of
energy, our $\sigma_{xx}$ spectrum
 is generally featureless,
with the exception of the minimum at 800-1000 meV - in remarkably
close agreement with experimental as well as model calculation
 results.\cite{hanki} In the inset, we show the effect of including
the Drude contribution to the spectra in an extended energy range:
as expected, only in the low energy range ($< \sim$ 1.5 eV) does the
Drude term significantly contribute to the total spectra. In
particular, the above mentioned minimum is present in both total and
interband-only terms, but it is much more enhanced upon inclusion of
the Drude term, therefore improving the agreement with experiment.

The real and imaginary parts of the dielectric constant,
$\epsilon_{xx}$\cite{Epsi}, are shown in Fig.~\ref{optical} (b) and
(c), respectively, and compared with experimental ellipsometry
measurements.\cite{burch} The agreement is rather good, as far as
the overall features are concerned. However, there is a sizeable
 discrepancy in the energy position
 of  peaks and valleys. This has probably to be ascribed
to the neglect of self--energy corrections or to the incorrect
treatment of correlation effects, resulting in an only partial
  description of the underlying electronic structure.
  In order to improve the agreement with experiment, we used the
so-called $\lambda$-fitting procedure, suggested by Rhee {\em et
al.} (see Ref.~\cite{lambda} for details), which is an
oversimplified approach to include a rescaling in the excitation
energy spectrum,
 avoiding the complicated task of properly evaluating  self-energy
effects.  Using a value of $\lambda \sim$ $-$0.1 eV,   both the real
and imaginary spectra were reasonably reproduced.

\section{$\vec{k}$ space dissection}\label{k-section}

In this section, we study the origin of the spectral features of Re
$\sigma_{xx}$. The qualitative analysis discussed in the previous
section is based on the band structure which highlights only
transitions involving electronic states at symmetry points or lines
in the irreducible BZ (IBZ). Here we examine the spectrum in further
detail, taking into account the contributions coming from the full
IBZ. Thus, we sample the BZ with an 8$\times$8$\times$8 cubic shell
and evaluate the matrix elements in Eq.~\ref{KUBO} in a  small
energy window ($\delta_{e}$=0.2 eV) centered  around the spectral
feature of interest. For each \textbf{k}-point, the contribution to
the conductivity ($\sigma_{\textbf{k}}^{\delta_{e}}$) was evaluated
calculating the dipole matrix elements between the initial and final
states contributing within the selected  energy window. Finally, the
$\sigma_{\textbf{k}}^{\delta_{e}}$ have been normalized to unity,
$\widetilde{\sigma}_{\textbf{k}}^{\delta_{e}}$. At each
\textbf{k}-point, the $\widetilde{\sigma}_{\textbf{k}}^{\delta_{e}}$
is graphically represented by an arrow whose length is proportional
to its magnitude. The results are shown in Fig.~\ref{IBZ}.

From this plot, one can easily visualize the region of
\textbf{k}-space contributing  to the features in
 the conductivity spectra.
 We focus on the maximum of Re $\sigma_{xx}$, at 3.8 eV.
 Figure~\ref{IBZ} shows the origin of all interband transitions in
 the range 3.6-4.0 eV, contributing to the maximum of Re
 $\sigma_{xx}$. First, we note that different regions in the IBZ
 contribute  to the conductivity very differently:
the largest contributions come from the region around the R point,
 and   the strength of the  contributions  decreases  upon moving towards M and X.
  The {\bf k}-points where large interband
  transitions occur belong to a line parallel to M-R, suggesting  that these are
  transitions between bands or band pairs which are very flat
  throughout much of this symmetry line or transitions with
  very large dipole matrix elements.
  The contributions from planes parallel to R-X-M  rapidly
  decrease in amplitude and become negligible for \textbf{k}-points
  close to $\Gamma$. From this, it is clear that the most active region
  in the IBZ in shaping the maximum of Re $\sigma_{xx}$ is the region parallel to the  R-X-M
  plane, near the edge of IBZ. 

\section{Magneto-optical properties}\label{optic-magneto}


We now turn our attention to the Kerr spectra. Equation~\ref{Kerr}
suggests that the Kerr angle could be enhanced by a large
magneto-optical component ($\sigma_{xy}$) and a small optical
component ($\sigma_{xx}$). To investigate the Kerr spectra on a
qualitative level, one can consider the separate contributions of
the numerator $\sigma_{xy}(\omega)$ and the denominator
$D(\omega)=\sigma_{xx}(\omega)\sqrt{1+\frac{4\pi}{\omega}\sigma_{xx}(\omega)}$.
The corresponding features in the imaginary part of the spectra are
then correlated to those observed in the Kerr spectra. This analysis
has been done in Ref.~\onlinecite{Silvia} by some of us, to which we
refer the reader for further details. Incidentally, we note that our
spectrum (presented in Fig.~\ref{numden}, see below) is quite
similar to those presented in Ref.~\cite{Silvia} for $x=6.25$\% and
$x=12.5$\%. We can conclude that the main features of the Kerr
spectra of (Ga,Mn)As do not depend to a great extent on the Mn
concentration, at least for the cases considered: this further
validates our choice to focus on the high concentration limit in
this study.

As pointed out by  Schoenes and Reim,~\cite{Schoenes}, the form
given by Eq.~\ref{Kerr} of the Kerr rotation is \emph{not} the most
suitable  to  discuss due to the complex denominator. Indeed, it is possibile to express the
real part of the left hand side of Eq.~\ref{Kerr} (that is the Kerr
angle) in terms of the real part of the \emph{complex fraction}
appearing in the right hand side of Eq.~\ref{Kerr}. In this way, one
ends up with a \emph{real fraction} where the denominator ("optical
component") depends \emph{only} on Re $\sigma_{xx}$ and Im
$\sigma_{xx}$ and the numerator ("magneto-optical component")
depends on \emph{both} the real and imaginary parts of $\sigma_{xx}$
\emph{and} $\sigma_{xy}$ (see the Appendix):

\begin{equation}\label{eqreale}
   \theta_{K}=-\frac{A(\omega)p(\omega)+B(\omega)q(\omega)}{D_{0}(\omega)D_{1}(\omega)}
\end{equation}
 with $A=R\sigma_{xy}R\sigma_{xx}-I\sigma_{xy}I\sigma_{xx}$
 and
$B=I\sigma_{xy}R\sigma_{xx}-R\sigma_{xy}I\sigma_{xx}$. (Here, 
R and I refer to Real and Imaginary part,
respectively).  For the definitions of $p$, $q$, $D_{0}$, and
$D_{1}$ we refer to the Appendix. We further define
N$_{1}=-A(\omega)p(\omega)$, N$_{2}=-B(\omega)q(\omega)$,
N=N$_{1}$+N$_{2}$ and D=D$_{0}$D$_{1}$ so that $\theta_{K}$ = N/D.
We will show that this decomposition can be useful to study the
correlations between  features of the Kerr spectra and 
electronic structure.

In Fig.~\ref{numden},  we show:  ellipticity
($\varepsilon_{K}$), and Kerr angle ($\theta_{K}$, in decimal
degree), in panel (a); N$_{1}$, N$_{2}$ and N in panel (b);
D$_{0}$, D$_{0}^{-1}$, and Re $\sigma_{xx}$
 (the spectrum has been rescaled in such a way
 that   Re $\sigma_{xx}$ and D$_{0}$ have the same maximum value) in panel (c); 
 D$_{1}$,
 and D$_{1}^{-1}$  in panel (d); and finally D, Re $\sigma_{xx}$
  (the spectrum has again been rescaled in such a way
 that   Re $\sigma_{xx}$ and D have the same maximum value), and
 D$^{-1}$, in panel (e).
 The notation $D_{...}^{-1}$
means $1/D_{...}$ (reciprocal function). All the quantities, except
the Kerr angle, are in arbitrary units.

Let us focus on the  Kerr ellipticity and rotation [panel (a)].
Clearly, they are related: when the Kerr ellipticity crosses the
zero line, a peak  appears in the Kerr rotation spectra  due to the
Kramers-Kroning relations.\cite{Weng} It is interesting to note that
the theoretical ellipticity curve crosses/touches the zero axis a
number of times, hence suggesting that the incident linearly
polarized light, at these frequencies, would stay as linearly
polarized only. The Kerr rotation is characterized by several spikes
as well as sign reversals in the whole energy window. In particular,
there is a first magneto-optical resonance (0.56$^\circ$)
 at $\hbar\omega\sim$ 0.2 eV and other main peaks are located at $\sim$
1, 2 and 6 eV. The arrows in the Kerr spectra mark the main features
(positive and negative). From Fig.~\ref{numden} [panels (a),(b)], we
see that the shape of the Kerr spectra is mainly determined by
N$_{1}$ in the low energy range (0-3 eV), by both N$_{1}$ and
N$_{2}$ in the medium energy range (3-7 eV), and by N$_{2}$ in the
high energy range (7-10 eV). Very interestingly, the overall
spectral trend of the Kerr angle, such as the peaks and sign
reversal positions, are very close to that of the numerator
N($\omega$) [panel (b)], and only the relative heights of the spikes
are different. As expected, the zero of N($\omega$) fixes the zero
in the Kerr spectra. We can draw the first conclusion: N($\omega$)
determines the overall trend of the Kerr spectra, and, in
particular, the presence of maxima and minima. We also stress that
the analytic form of the numerator of the Kerr angle (see
Eq.~\ref{eqreale} and the Appendix) \emph{entangles} in a rather
complicated way \emph{all} the real and imaginary components of the
conductivity tensor: there is no simple guideline to link the trend
of a part of the Kerr spectra to the spectral trend of a specific
(real or imaginary) tensor component in the same energy range.

We now focus on the denominator D$^{-1}$. First we write
$\theta_{K}$=N($\omega$)D$^{-1}$($\omega$), so that D$^{-1}$
plays the role of an \emph{enhancement} factor. It is evident
that  both D$_{0}$ and D$_{1}$ are positive definite, that is they
never cross the zero line: the \emph{enhancement} factor does not
have any "resonant" behavior. In Fig.~\ref{numden} panel (c), we can
see that, remarkably, the spectral trend of D$_{0}$ is very
\emph{similar} to Re $\sigma_{xx}$. We might therefore infer that
D$^{-1}_{0}$ has large values whenever Re $\sigma_{xx}$ has its
smallest ones. In our case, this happens when the photon energy
approaches $\sim$ 0 eV. On the other hand, D$_{1}$ [panel (d)] does
not show any remarkable features, apart from the maximum at very low
energy. Considering that the reciprocal function D$^{-1}_{0}$
[panel (c)] exhibits a maximum at $\sim$ 0 eV and strongly decreases
with increasing energy, D$^{-1}_{1}$ has a minimum at $\sim$ 0 eV and
monotonically increases with energy. D$^{-1}_{1}$ is also
featureless, in almost the whole energy window.
From panel (e), we see that D$^{-1}$ has a large peak around $\sim$
0.5 eV, then decreases very fast to an almost constant value, and
slowly increases at energies larger than $\sim$ 5 eV. Clearly, the
shoulder of D$^{-1}$ at low energy comes from the behavior of
D$^{-1}_{0}$ and D$^{-1}_{1}$ in the same energy range, whereas the
increase of D$^{-1}$ at high energy is mainly due to the behavior of
D$^{-1}_{1}$   in the corresponding energy range.

These results clearly explain the \emph{origin} of the Kerr spike
(0.56$^\circ$) at $\sim$ 0.2 eV. The numerator  determines the
existence and location of the spikes: whether they give rise to a
strong Kerr angle follows entirely from the \emph{enhancement}
factor, D$^{-1}$, which, in this case, magnifies the peaks at low
energy and suppresses those at higher
 energy. The importance of D$^{-1}$ is highlighted by again considering
N($\omega$): we would expect large Kerr angle values at higher
energy (around 6 eV). However, the corresponding peaks are
 suppressed by the D$^{-1}$ factor. Hence, we are ready to conclude that the
 large Kerr peak at low  energy has an ``optical" origin; the features
 at high  energy have an MO origin (they follow the trend of the
 numerator) but they do not  result in large peaks due to
 the damped behavior of $D^{-1}$ in this same energy range.
  This confirms in a more transparent and direct way what was already observed
 in Ref.~\cite{Silvia} More important, the  similar trend of
 $D_{0}(\omega)$ (D)  and $R\sigma_{xx}$ suggests that the presence of minima
 in the $R\sigma_{xx}$ spectra may give rise to strong Kerr
 angles in the same energy range. On the other hand, $R\sigma_{xx}$ is directly linked
 to the optical transitions, that is to details of the electronic structure of the
 materials: this could have far reaching consequences in
 \emph{magneto-optical Kerr effect engineering}, provided that the
 electronic structure of the material can be tuned correctly in
 such a way as to \emph{enhance} the  features of the
 Re $\sigma_{xx}$ minima.

\section{Conclusion}\label{conclu}

We presented results of first-principles calculations of the
magneto-optical properties of (Ga,Mn)As  within density functional
theory aimed at investigating in great detail the role of the
electronic and magnetic properties in determining the magnetooptical
behaviour of the material.
 The spectral features of the optical
tensor in the 0-10 eV energy range were analyzed in terms of the
band structure and density of states and the role of the dipole
matrix elements was highlighted in terms of Brillouin zone
\emph{dissection}.

We found that  different types of interband transitions contribute
in shaping the conductivity tensor. The dipole matrix elements play
a key role greatly affecting the optical spectra in the low as well
as high energy ranges. Brillouin zone \emph{dissection} reveals that
different regions in the Brillouin zone (not necessarily symmetry
points or lines) can contribute very differently in shaping the
conductivity tensor; we find that the most active region in the IBZ
is around the R point.

Moreover, since the Kerr rotation spectrum is a result of a complex
entanglement of real and imaginary contrbutions, we presented a
possible way to analyze the calculated Kerr spectra in terms of a
real representation of the Kerr angle. To the best of our knowledge,
this has not  been used as an analysis tool in the past, although it
is implicit in Eq.~\ref{Kerr}. Using this representation, we have
clearly elucidated the origin of the features of the Kerr spectra
and have highlighted the role of the minima of Re $\sigma_{xx}$
which can \emph{possibly} correlate with a large Kerr angle (in our
case, at low energies). Nevertheless, the results given above
support the conclusion that at present it is difficult to give
simple rules for the occurrence of large spectral features at
specific laser-light frequencies. Indeed, both spin-orbit and
exchange splitting effects are \emph{entangled} in $N(\omega)$, which
determines the overall behavior of the Kerr spectra and eventually
the strength of the Kerr features. No simple rules can be given to
\emph{disentangle} their effects on the absolute magnitude of the
Kerr angle. On the other hand, we found that part of the Kerr
spectrum (in our case, the low energy range) is very sensitive to
the shape of Re $\sigma_{xx}$, which is very closely related to
details of the band structure of the materials.

While it is obvious that without a precise
 knowledge of the material band structure it is not possible to make a priori
predictions of the Kerr angle magnitude,  it is also true that a
proper tuning of the band structure would allow one to consequently
tune the magnitude and location of spectral peaks through the
D$_{0}$ term, whose spectral behavior is very close to Re
$\sigma_{xx}$. However, further studies are needed in order to
better elucidate how this proper tuning of the underlying electronic
structure can be achieved (i.e. by making use of alloying,
pressure, strain field) in order to enhance the minima of Re
$\sigma_{xx}$. Therefore, we expect that computational materials
design can substantially contribute to \emph{magneto-optical Kerr
effect engineering} in the future.

\section{Acknowledgments}
One of the authors (A. Stroppa) thanks G. Kresse for useful comments
and fruitful discussions; this work started during a stay at
Dipartimento di Fisica Teorica, Universit\`a degli Studi di Trieste,
Strada Costiera 11, I-34014 Trieste, Italy and INFM-CNR DEMOCRITOS
National Simulation Center, Trieste, Italy.

Work at Northwestern University was supported by the United States
NSF (through its MRSEC program at the Materials Research Center).

\begin{flushleft}\label{Analysis}
    {\bf APPENDIX}
  \end{flushleft}
Here we give an  explicit  representation of the \emph{complex} Kerr
angle ($\theta_{k}+i\eta_{k}$, Eq.~\ref{Kerr}) in the form
$\alpha$+$i\beta$, where $\alpha$ and $\beta$ are real functions:
the Kerr angle ($\theta_{k}$) corresponds to $\alpha$. To this end,
we need to calculate the square root of the complex number in the
denominator of Eq.~\ref{Kerr}. It can be helpful to make use of the
following formula\cite{squareroot}: given $\sqrt{a+ib}=p+iq$ we have

\begin{equation}\label{root1}
p=\frac{1}{\sqrt{2}}\sqrt{\sqrt{a^{2}+b^{2}}+a}\hspace{1cm}
q=\frac{sgn(b)}{\sqrt{2}}\sqrt{\sqrt{a^{2}+b^{2}}-a}
\end{equation}
where $sgn(b)$ is the sign of $b$. In our case,
$$\sqrt{1+\frac{4\pi i}{\omega}\sigma_{xx}}=\sqrt{(1-\frac{4\pi}{\omega}I\sigma_{xx})+i\frac{4\pi}{\omega}R\sigma_{xx}}$$
where R and I stand for real and imaginary part and
$a=1-\frac{4\pi}{\omega}I\sigma_{xx}$,
$b=\frac{4\pi}{\omega}R\sigma_{xx}$. By definition, $b$ is positive.

In order to get rid of the complex number in the denominator of
Eq.~\ref{Kerr}, we multiply numerator and denominator   by
$\sigma_{xx}^{*}(p-iq)$.  After some algebra, we
 express:
\begin{equation}\label{newkerr}
\theta_{k}+i\eta_{k}=-\frac{(Ap+Bq)+i(Bp-Aq)}{D_{0}D_{1}}
\end{equation}
Then the Kerr angle is:
\begin{equation}\label{newkerr1}
\theta_{k}=-\frac{(Ap+Bq)}{D_{0}D_{1}}
\end{equation}
and the ellipticity is:
\begin{equation}\label{newellipt}
\eta_{k}=\frac{Aq -Bp}{D_{0}D_{1}}
\end{equation}
where $A=R\sigma_{xy}R\sigma_{xx}+I\sigma_{xy}I\sigma_{xx}$,
$B=I\sigma_{xy}R\sigma_{xx}-R\sigma_{xy}I\sigma_{xx}$,
$D_{0}=(R\sigma_{xx})^{2}+(I\sigma_{xx})^{2}$ and
$D_{1}=\sqrt{(1-\frac{8\pi}{\omega})I\sigma_{xx}+\frac{16\pi^{2}}{\omega^{2}}D_{0}}$

 In this representation, an explicit analytic relation for
the Kerr spike can be derived using the Kramers-Kroning
transformation: whenever the ellipticity vanishes, a spike in the
Kerr spectra appears.\cite{Weng} Thus, imposing that $\eta_{k}=0$,
we derive an analytic relation for $\theta_{k}$ which holds for the
maximum or minimum in the Kerr spectra corresponding to zero
ellipticity. From $\eta_{k}=0$, we have either (a) $A=\frac{Bp}{q}$
or (b) $B=\frac{Aq}{p}$ and using Eq.~\ref{newkerr1} we obtain:
\begin{equation}
\theta_{k}=-A\frac{p^{2}+q^{2}}{pD_{0}D_{1}}=-B\frac{p^{2}+q^{2}}{qD_{0}D_{1}}
\end{equation}
A necessary condition to produce a Kerr spike is that
$\frac{A}{B}=\frac{p}{q}$.

\newpage

\begin{figure}[!hbp]
\caption{Total and atomic- angular momentum- projected density of
states (DOS) in   Ga$_{.75}$Mn$_{.25}$As. Positive and negative
curves refer to spin-up and spin-down components, respectively. The
Fermi level is fixed at zero energy.}\label{dos}
\end{figure}

\begin{figure}
\caption{(Color on line) Spin-polarized band structure for
Ga$_{.75}$Mn$_{.25}$GAs along the symmetry lines
$\Gamma$-X-M-$\Gamma$-R-M in the irreducible BZ plotted without
spin-orbit interaction for clarity: parts (a) and (b) refer to
spin-up (-down), respectively. The horizontal line marks the Fermi
level (zero energy). Numbers in round brackets at $\Gamma$ label
groups of bands  discussed in the text. The vertical  lines indicate
the different possible band to band transitions at 0.4, 1.1, 2.8,
3.8 eV within 0.1 eV (solid black, dotted blue, dashed red and
dot-dashed orange respectively). The labels for symmetry points and
lines are according to the standard group-theoretical notation of
Bouckaert, Smoluchowski, and Wigner. Spin-orbit effects are
neglected.}\label{bands}
\end{figure}

\begin{figure}
\caption{Calculated real (a) and imaginary (b) parts of diagonal and
off-diagonal components of the optical conductivity. The Drude term
has not been included in the spectra. Vertical lines in Re
$\sigma_{xx}$ indicate features discussed in the
text.}\label{sigmaxx}
\end{figure}

\begin{figure}
\caption{Selected contributions from spin-up (solid line) and
spin-down (dashed line) transitions to the total  Re $\sigma_{xx}$
spectra. The joint density of states (JDOS) is  also reported, in
arbitrary units.}\label{joint}
\end{figure}

\begin{figure}
\caption{(a) Real part of the diagonal conductivity: our calculated
spectrum (bold line) is compared with the results of model
calculations (dashed line) from Ref.\protect\cite{hanki} and
experiments (dotted line) from Ref.\protect\cite{singley}. The inset
in panel (a) shows the difference between the interband and the
total (interband + intraband) conductivity. Panels (b) and (c) show
the calculated real and imaginary parts of the dielectric constant,
respectively, with (bold line) and without (dashed line)  $\lambda$
fitting. Ellipsometry results from Ref.\protect\cite{burch} are also
shown (solid line).} \label{optical}
\end{figure}

\begin{figure}
\caption{Prospective view of the irreducible wedge of Brillouin zone
showing the origin of the interband transitions at $\sim$ 3.8 eV.
The length of the arrows shows the relative  contributions to Re
$\sigma_{xx}$ in the energy interval 3.6-4.0 eV (see
Fig.~\ref{sigmaxx}). Increasing of the length of the arrows
represents an increasing contribution. See text for further
details.}\label{IBZ}
\end{figure}

\begin{figure}
\caption{(Color online) (a) Ellipticity and  Kerr angle; the main
features of the Kerr spectra are indicated by arrows; (b) Different
contributions to the numerator of the formula presented in
Sect.~\ref{optic-magneto} and the Appendix (arrows indicate the main
spectral features);
 (c,d,e) decomposition of the denominator (see text for
details). Arbitrary units are used except for the Kerr angle (in
decimal degree).  See text for details.}\label{numden}
\end{figure}

\clearpage
\includegraphics[scale=.7,angle=-90]{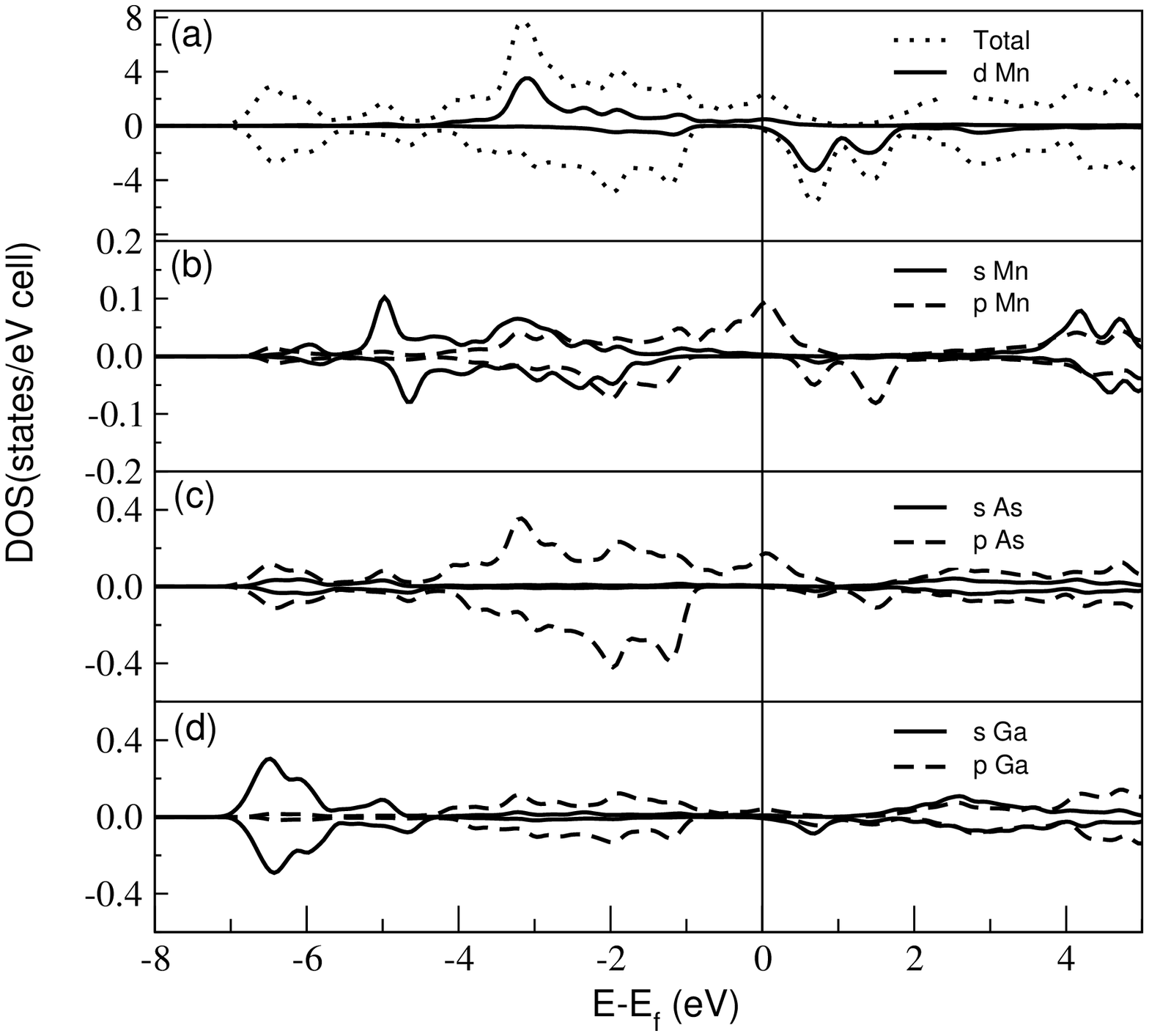}
\\Fig. 1

\clearpage
\includegraphics[scale=.7,angle=0]{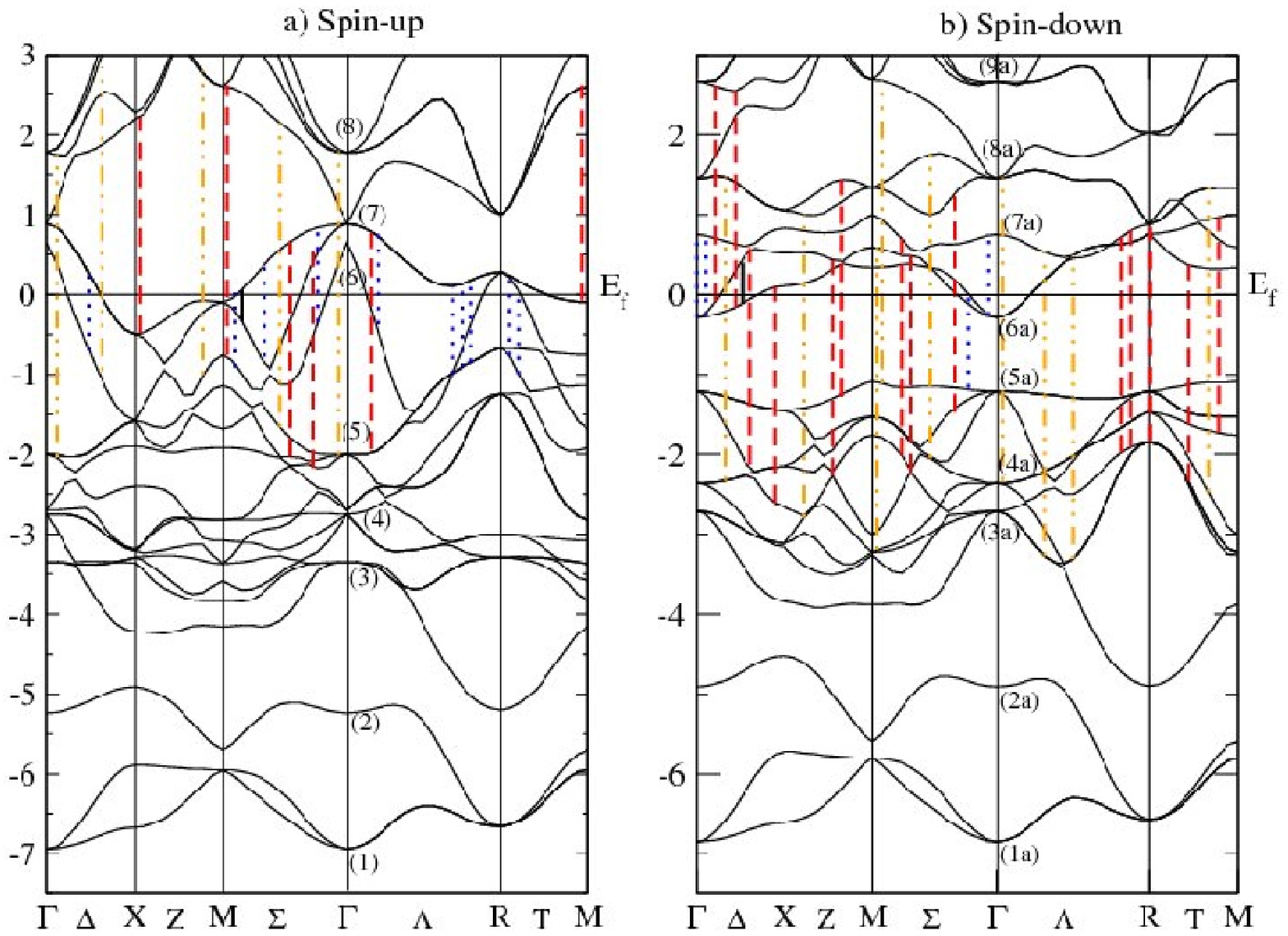}
\\Fig. 2

\clearpage
\includegraphics[scale=.7,angle=-90]{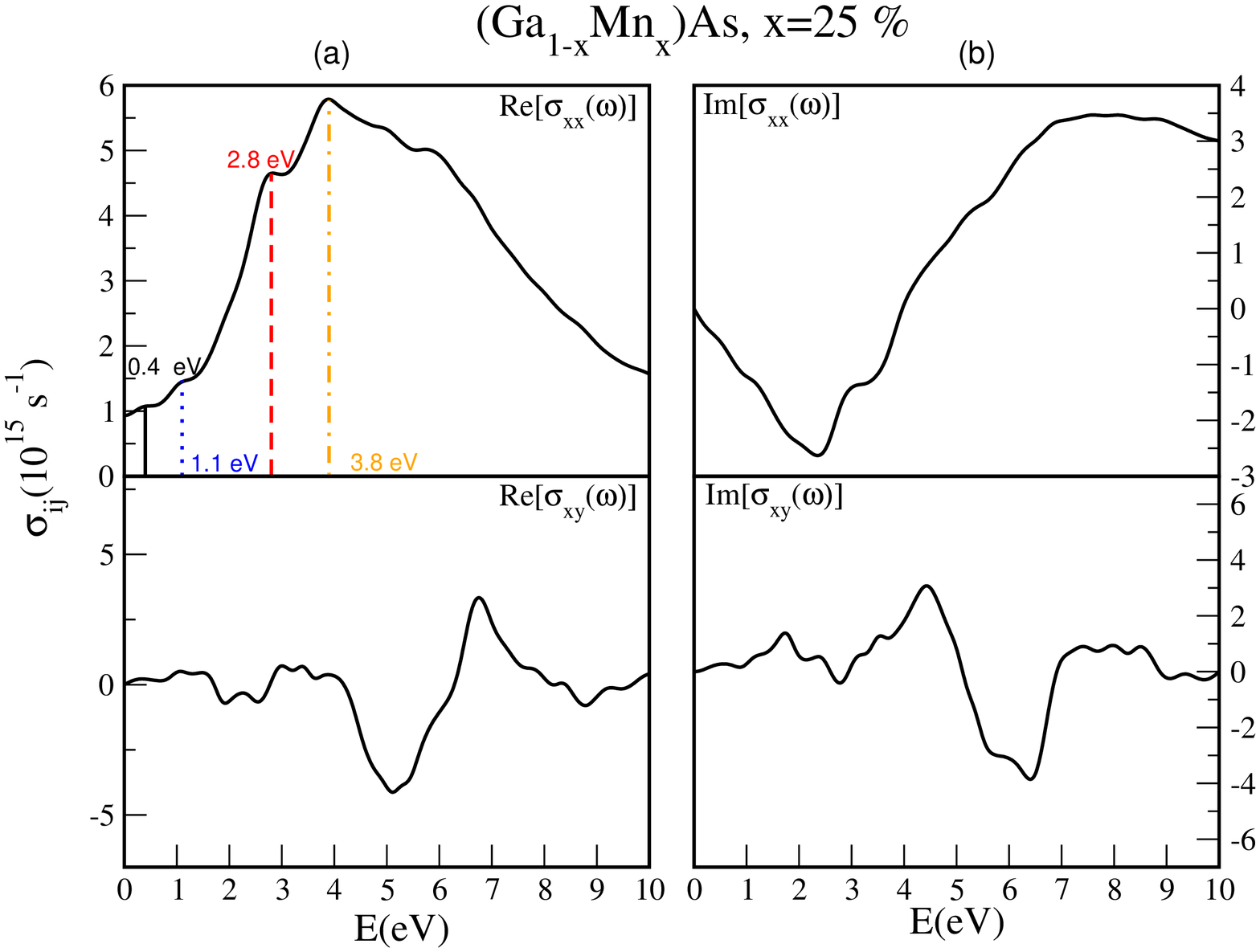}
\\Fig. 3

\clearpage
\includegraphics[scale=.7,angle=-90]{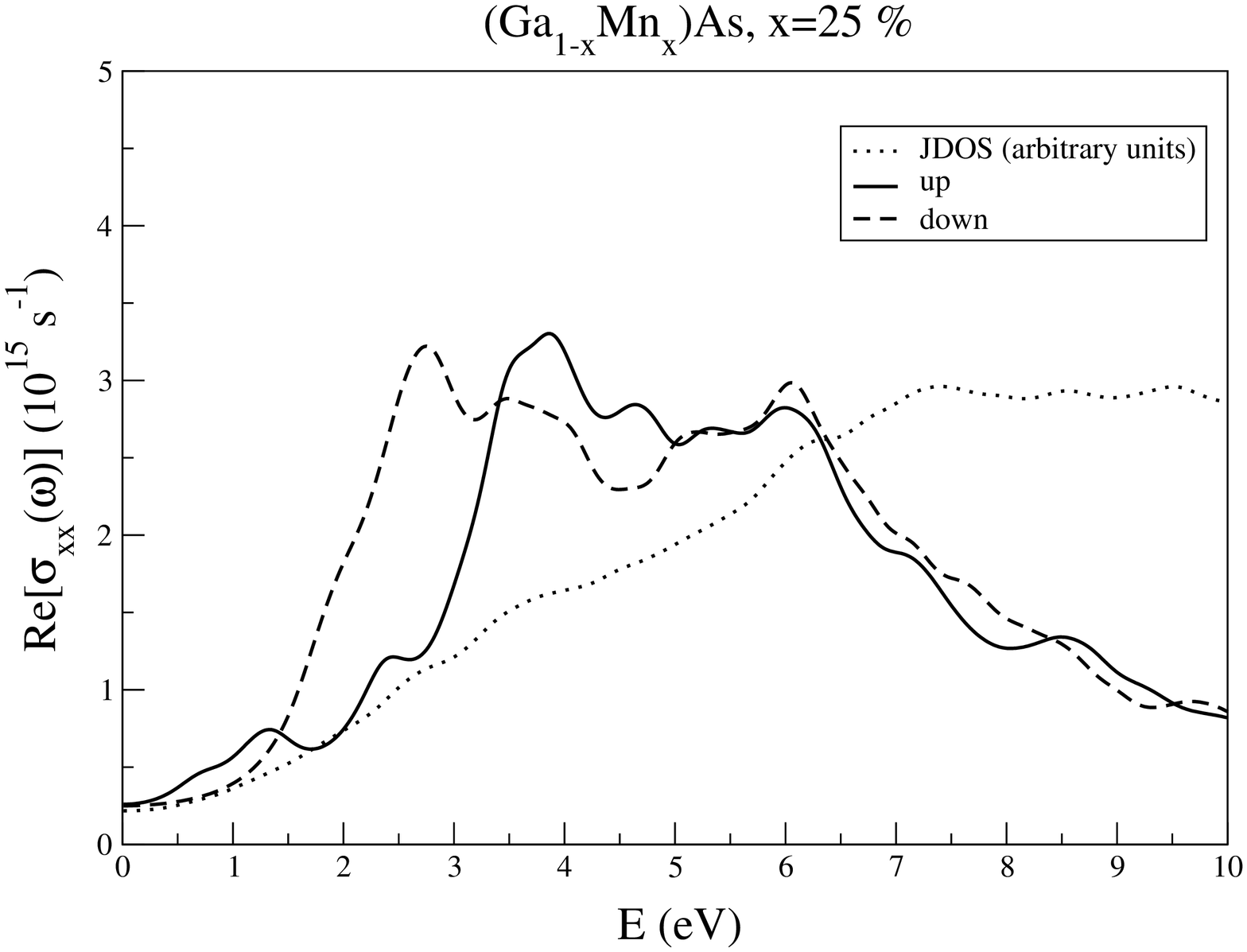}
\\Fig. 4

\clearpage
\includegraphics[scale=.7,angle=0]{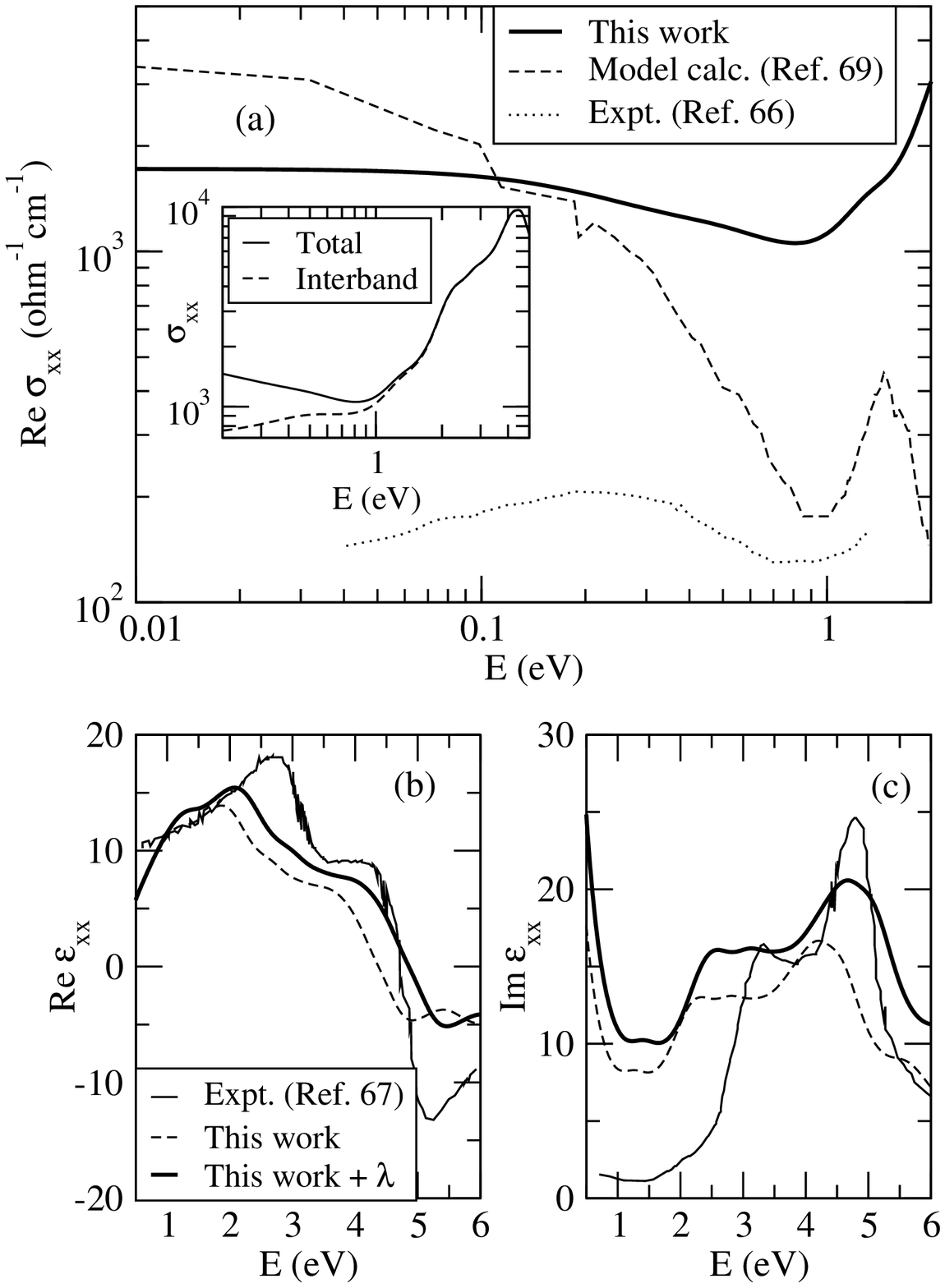}
\\Fig. 5

\clearpage
\includegraphics[scale=.4,angle=0]{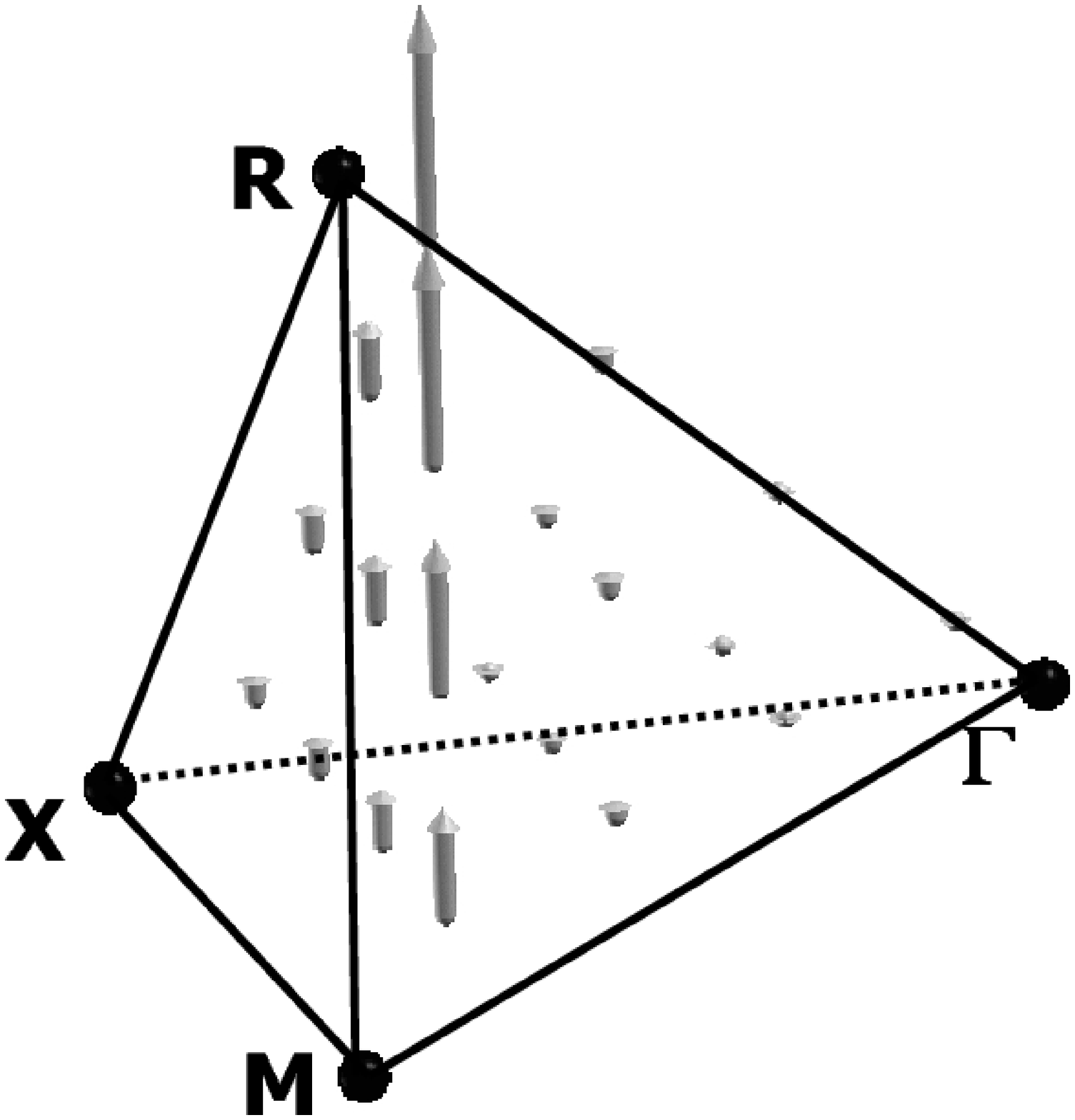}
\\Fig. 6

\clearpage
\includegraphics[scale=.7,angle=-90]{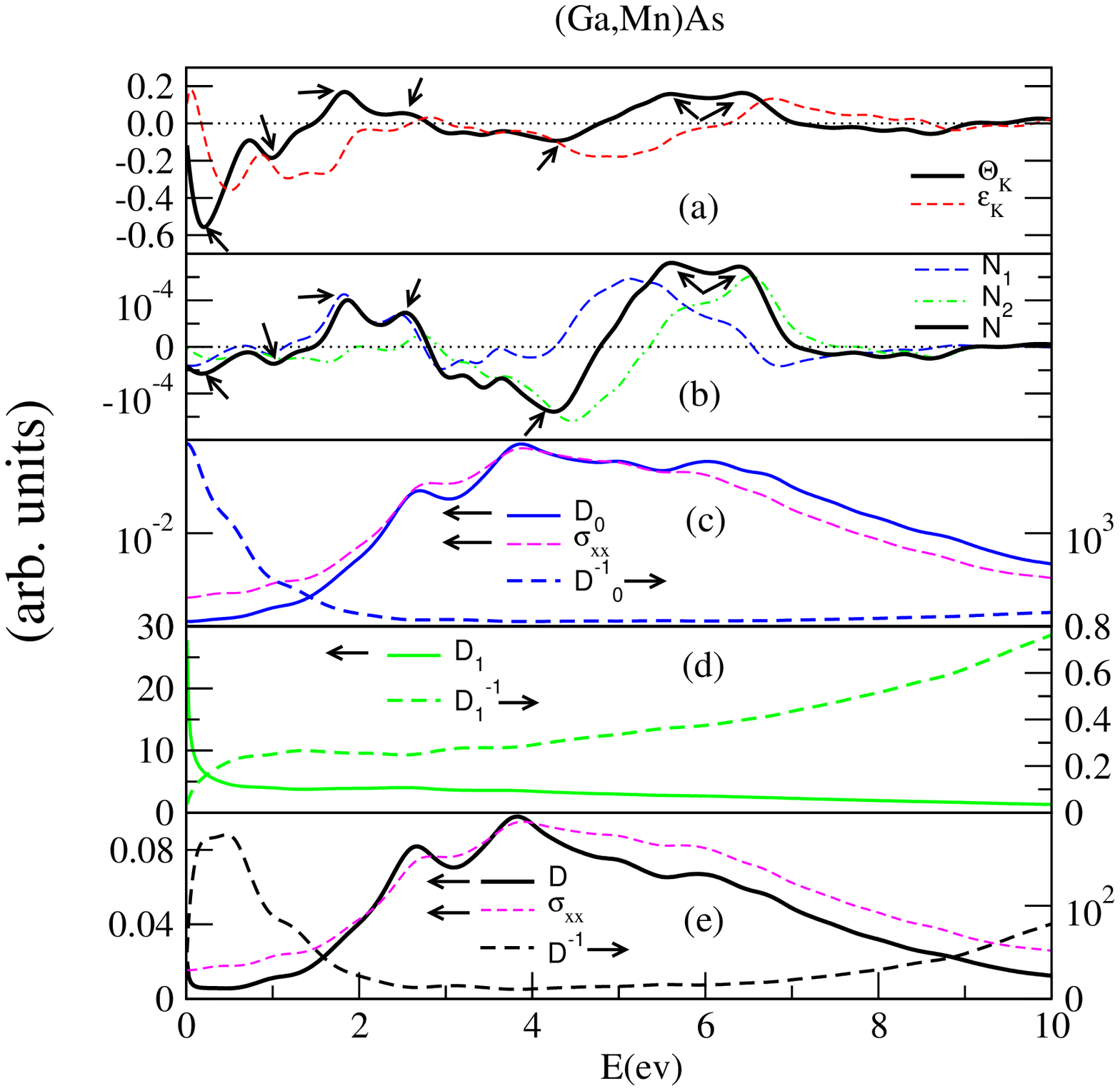}
\\Fig. 7


\begin{thebibliography}{99}
\bibitem{Spintransition} Spin-flip transitions are allowed within
the electric dipole approximation when spin-orbit coupling is
included, but are omitted in our work since their effect is
negligible.
\end{thebibliography}
\end{document}